\documentclass[aps, reprint, groupedaddress, longbibliography]{revtex4-2}
\usepackage{graphicx, amsmath, mhchem}
\usepackage{subfigure, amssymb}
\usepackage{float} 
\usepackage[makeroom]{cancel}
\usepackage{color}

\graphicspath{ {Figures_arch/} }

\begin{document}

\title{Kink propagation in the Artificial Axon}
\author{Xinyi Qi}
\author{Giovanni Zocchi}
\email{zocchi@physics.ucla.edu}
\affiliation{Department of Physics and Astronomy, University of California - Los Angeles}

\begin{abstract}

\noindent The Artificial Axon is a unique synthetic system, based on biomolecular components, which supports 
action potentials. Here we consider, theoretically, the corresponding space extended system, and discuss 
the occurrence of solitary waves, or kinks. In contrast to action potentials, stationary kinks are possible. 
We point out an analogy with the interface separating 
two condensed matter phases, though our kinks are always non-equilibrium, dissipative structures, 
even when stationary.

\end{abstract}

\maketitle

\noindent {\it Introduction.} The Artificial Axon (AA) is a synthetic structure designed 
to support action potentials, thus generating these excitations for the first time outside the living cell. 
The system is based on 
the same microscopic mechanism as that operating in neurons, the basic components being: a phospholipid 
bilayer with embedded voltage gated ion channels, and an ionic gradient as the energy source. 
However, while a real axon has at least two ion channel species and 
opposite ionic gradients across the cell membrane, the AA has only one. In the experiments, 
a current limited voltage clamp (CLVC) takes the role of a second ionic gradient \cite{Amila2016, Hector2017}. 
The experimental system in \cite{Hector2017} is built around a 
$\sim 100 \, \mu m$ size black lipid membrane. As a dynamical system for the voltage, it operates 
in zero space dimensions (similar to the "space clamp" setup with real axons \cite{Marmont1949, Koch_book}). 
That is, each side of the membrane is basically an  equi-potential surface (the name 
Artificial Axon, while a misnomer in this respect, is historical \cite{Amila2016} and we propose to keep it 
for the original and future versions). Inspired by this system, here we consider - theoretically - the 
corresponding space extended dynamical system. We focus on the existence of solitary wave solutions, 
or propagating kinks (we will use the two terms interchangeably, to mean a front which propagates 
keeping its shape). Kinks appear in many areas of condensed matter physics \cite{Chaikin_Lub_Book}, 
from domain walls in magnetic materials \cite{Buijnsters2014, Kolar1996} to pattern forming 
chemical reactions \cite{Rotermund1991}. 
Our particular nonlinear structures come from a dissection, so to speak, of the mechanism of action potential  
generation. We show the existence of travelling kinks in our system, and study numerically 
their characteristics in relation to the control parameters, which are the command voltage and the 
conductance of the CLVC. Then we discuss 
a "normal form" for this class of dynamical systems, highlighting the relation with other kinks separating 
two condensed matter phases, such as the nematic - isotropic interface in liquid crystals. 
The nonlinearities which thus arise retrace the development of simplified models of the Hodgkin-Huxley 
axon \cite{hodgkin_quantitative_1952}, such as introduced 60 years ago by Fitzhugh \cite{Fitzhugh} 
and Nagumo et al \cite{Nagumo1962}. Looking at kinks thus provides a somewhat different perspective 
on a classic topic in the study of excitable media. \\

\noindent {\it Results.}
We consider the AA in one space dimension. The physical system we have in mind is a $\sim 1 \, cm$ long, 
$\sim 100 \, \mu m$ wide supported strip of lipid bilayer with one species of voltage gated ion channels 
embedded. The bilayer might be anchored to the solid surface so as to leave a sub-micron gap (the "inside" 
of the axon) in between. At present, the stability of the bilayer stands in the way of a practical realization, 
but this problem is not unsurmountable. The bilayer acting essentially like the dielectric in a parallel plates 
capacitance, the local charge density is related to the voltage by $(\partial / \partial t) \rho (x, t) = 
c \, (\partial / \partial t) V(x, t)$ where $c$ and $\rho$ are capacitance and charge per unit length, respectively. 
The current inside the axon follows Ohm's law: $j = - (1/r) (\partial V / \partial x)$ where $r$ is the resistance 
per unit length; then charge conservation leads to the diffusion equation for the potential: 
$(\partial V(x, t) / \partial t) - (1/(r c)) (\partial^2 V(x, t) / \partial x^2) = 0$ . In the AA, an ionic gradient 
(of $K^+$ ions) across the membrane leads to an equilibrium (Nernst) potential 
$V_N = (T / |e|) \, ln ([K^+]_{out} / [K^+]_{in})$ , but the system is held off equilibrium by the current 
injected through a current limited voltage clamp (CLVC) \cite{Amila2016}. The active elements are voltage gated 
potassium channels inserted in the membrane: these are molecular pores which, in the open state, 
selectively conduct $K^+$ ions. The KvAP channel used in \cite{Hector2017, Hector2019} has three 
functionally distinct states: open, closed, and inactive; the presence of the inactive state allows the system 
to generate action potentials. Here we consider the simpler case of a "fast" channel with no inactivation. 
Then the channels can be described by an equilibrium function $P_O (V)$ which gives the probability that 
the channel is open if the local voltage is $V$. Introducing the current sources in the diffusion equation 
above one arrives at the following $(1+1) D$ dynamical system: 

\begin{equation}
\begin{split}
\frac{\partial V(x, t)}{\partial t} - \frac{1}{r c} \frac{\partial^2 V}{\partial x^2} \, = \, 
 \frac{\chi}{c} P_O (V) [V_N - V(x, t)] \\ 
+ \frac{\chi_c}{c} [V_c - V(x, t)] 
\end{split}
\label{eq: DS1}
\end{equation}

\noindent V is the voltage inside the axon (referred to the grounded outside), and we assume a distributed 
"space clamp" for the CLVC (this would be provided by an electrode along the axon). Eq. (\ref{eq: DS1}) 
is of the general form of a reaction - diffusion system; these are usually studied in the context of pattern 
forming chemical reactions. For us it represents a continuum limit, i.e. we consider a uniform,  
distributed channel conductance instead of discrete, 
point-like ion channels. This is a mean field approximation which neglects correlations between nearby 
channels. The first term on the RHS of (\ref{eq: DS1}), when 
multiplied by $c$, is the channel current, proportional to the driving force $(V_N - V)$ ; 
$V_N$ is the Nernst potential, $\chi$ the conductance 
(per unit length) with channels open (i.e. $\chi = n \chi_0$ , $\chi_0$ single channel conductance, 
$n$ number of channels per unit length). The second term is the current injected by the clamp; 
$V_c$ is the clamp voltage (which is a control parameter in the experiments), $\chi_c$ the clamp conductance 
(per unit length), which is a second control parameter. The function $P_O (V)$ is a Fermi - Dirac distribution: 

\begin{equation}
P_O (V) \, = \, \frac{1}{exp[- q (V - V_0) / T] + 1}
\label{eq: FD}
\end{equation}

\noindent where $q$ is an effective (positive) gating charge and $V_0$ the midpoint voltage where 
$P_O (V_0) = 1/2$. To fix ideas, we will use parameters consistent with the AA in \cite{Hector2019} : \\ 
$V_N = 50 \, mV$ , $\chi / c = 100 \, s^{-1}$ , $\chi_c / c = 5 \, s^{-1}$ , $(1/rc) = 1 \, cm^2 / s$ , 
$V_0 = - 10 \, mV$ , $q/T = 0.08 \, (mV)^{-1}$. We use Gaussian units except that we express voltages 
in $mV$ : this is more convenient to relate to experimental systems. Also, the temperature 
in (\ref{eq: FD}) and elsewhere is in energy units; thus at room temperature $T/|e| \approx 0.025 \, mV$ 
where $e$ is the charge of the electron.\\ 

\begin{figure}
\includegraphics[width=\linewidth]{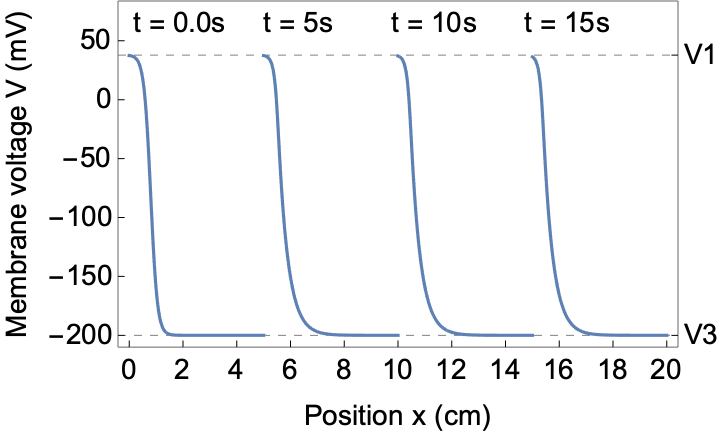}
\caption{The traveling kink solution $V(x, t)$ for (\ref{eq: DS1}), (\ref{eq: FD}). The plot shows snapshots of the kink 
at different times; the initial condition ($t = 0$) is a hyperbolic tangent. Parameter values are those given in 
the text, with a clamp voltage $V_c = - 200 \, mV$. The dotted horizontal lines show the fixed points $V_1$ 
and $V_3$. Notice that the shape of the kink shifts from the initial condition at t = 0.0s 
to a stable shape afterwards.}
\label{fig:kink_1}
\end{figure}

\begin{figure}
\includegraphics[width=\linewidth]{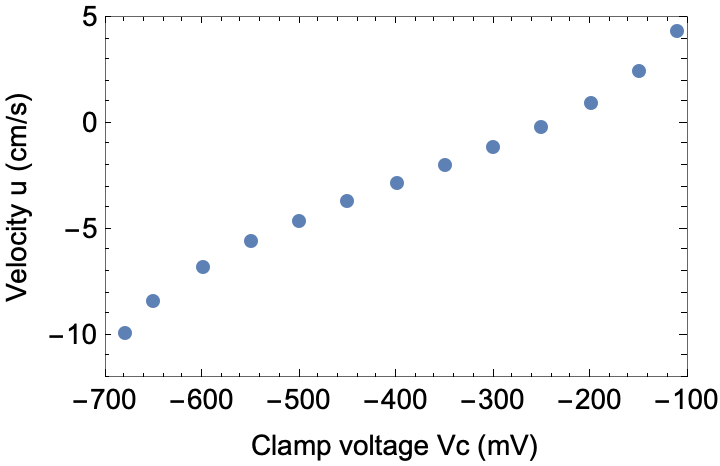}
\caption{Plot of kink velocity vs clamp voltage.  Parameter values are those given in the text. 
The velocity is determined by tracking the minimum 
of the ﬁrst derivative of $V(x, t)$, which corresponds to the inﬂection point of the kink-shaped wave front. 
The left most and right most data points are close to the values of $V_c$ beyond which the kink solution 
disappears. The graph is asymmetric with respect to right moving and left moving kinks. }
\label{fig:Vel_1}
\end{figure}

\noindent The possibility of travelling kink solutions of (\ref{eq: DS1}) and (\ref{eq: FD}) arises because, 
with the clamp 
at a negative voltage, say $V_c = - 100 \, mV$, there is a fixed point of (\ref{eq: DS1}) (a uniform, 
time independent solution) with $V(x, t) \approx V_N$ and open channels ($P_O (V) \approx 1$), 
namely $V = V_1 \approx (\chi V_N + \chi_c V_c) / (\chi + \chi_c)$. A second fixed point is 
$V(x, t) = V_3 \approx V_c$ and channels closed ($P_O (V) \approx 0$). A stable kink solution exists, 
asymptotically connecting these two stable fixed points (a third fixed point is unstable and will be discussed later).  
The essential parameters in (\ref{eq: DS1}) are the diffusion constant $D \equiv 1 / (rc)$ and $\chi /c$ ; 
from these we can form a characteristic length scale $\Delta = 1 / \sqrt{r \chi}$ which gives the scale 
of the width of the kink solution, and a characteristc velocity $v = D / \Delta = (1/c) \sqrt{\chi / r}$ 
which similarly gives the scale for the kink velocity. With the parameters above, $\Delta \approx 1 \, mm$ 
and $v \approx 10 \, cm/s$. Fig. \ref{fig:kink_1} shows snapshots of a travelling kink obtained by integrating 
(\ref{eq: DS1}) , (\ref{eq: FD}) using the parameters above and $V_c = - 200 \, mV$. The kink was launched 
with a hyperbolic tangent initial condition ($t = 0$ trace in Fig. \ref{fig:kink_1}); it is found to quickly 
(on a time scale $\sim c / \chi$) attain a stable limiting shape and thereafter travel at constant velocity. 
The velocity depends on the clamp voltage $V_c$, as shown in Fig. \ref{fig:Vel_1}. We measure it 
by tracking the inflection point of the solution $V(x, t)$. 
The solitary wave solution exists only for 
$V_c$ within certain bounds; correspondingly there is a maximum velocity of the kink, while the minimum 
velocity is zero, as we show below. 

Let us now analyze these solitary wave solutions (see e.g. \cite{Chaikin_Lub_Book}). 
Eq. (\ref{eq: DS1}) is of the form: 

\begin{equation}
\frac{\partial V(x, t)}{\partial t} - \frac{\partial^2 V}{\partial x^2} \, = \, g(V) 
\label{eq: DS2}
\end{equation}

\noindent where we have changed to non-dimensional variables using $\Delta = 1 / \sqrt{r \chi}$ , 
$\tau = c / \chi$ , $V_N$ as the units of length, time, and potential, respectively. Then, 

\begin{equation}
\begin{cases}
g(V) \, = \, P_O (V) [1 - V] + \frac{\chi_c}{\chi} \left [\frac{V_c}{V_N} - V \right ] \\ 
\\
P_O(V) \, = \, \left \{ exp [- \frac{q V_N}{T} (V - \frac{V_0}{V_N})] +1 \right \}^{-1}
\end{cases}
\label{eq: g1}
\end{equation}

\begin{figure}[H]
\includegraphics[width=\linewidth]{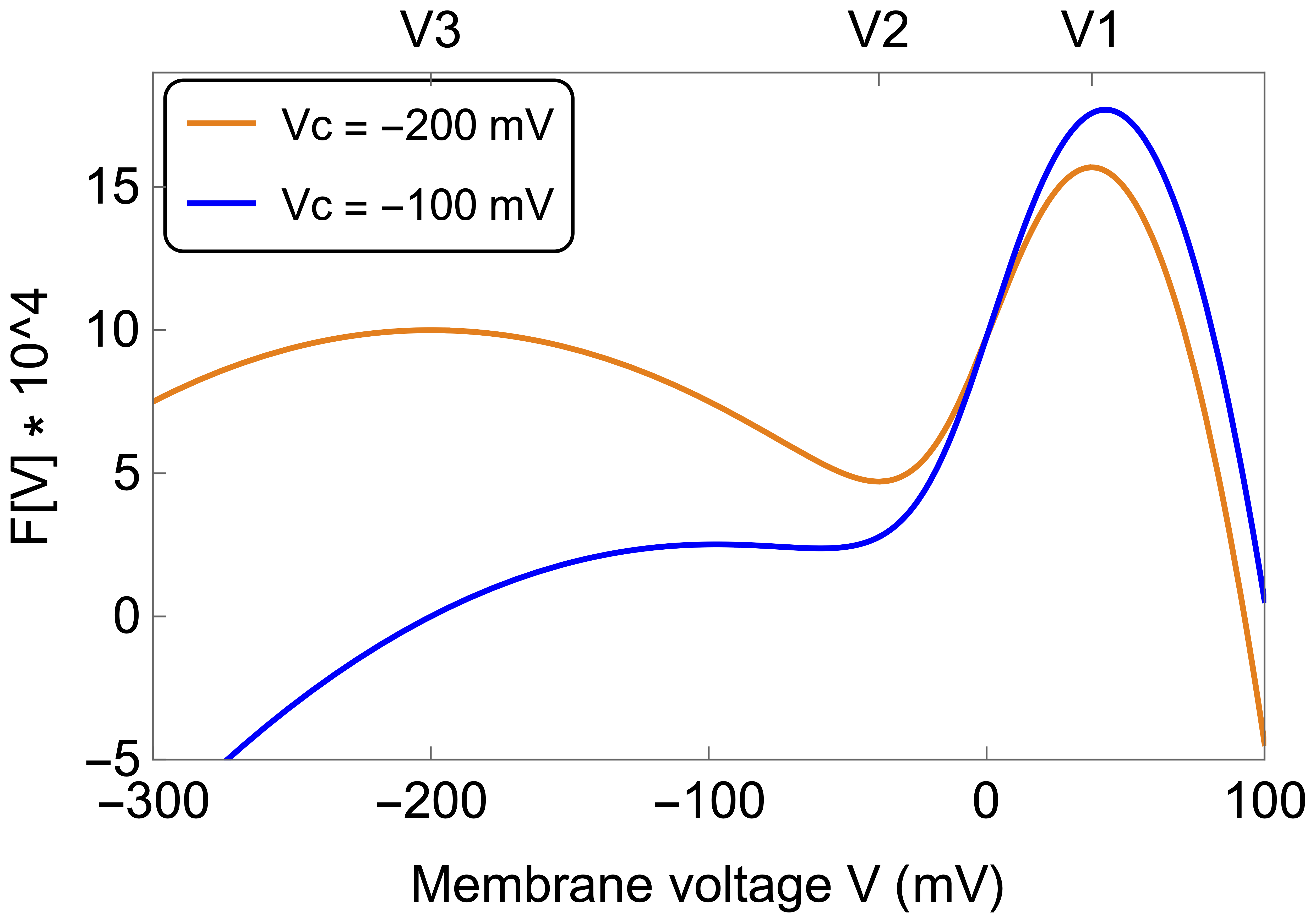}
\caption{The function $F (\varphi)$ obtained from (\ref{eq: g1}) vs the (dimensional) membrane voltage,  
for clamp voltages of  $-100 \, mV$ and $-200 \, mV$. Parameters are as given in the text.  
The fixed points $V_1$, $V_2$, $V_3$ shown refer to the yellow ($V_C = - 200 \, mV$) curve. 
As $V_C$ is decreased below $- 200 \, mV$ the global maximum becomes the secondary maximum and 
vice-versa. Increasing $V_c$ above $- 100 \, mV$, the secondary maximum eventually disappears, 
at which point there is no kink solution. }
\label{fig:F_vs_phi_1}
\end{figure}

\noindent We look for a travelling wave solution: $V(x, t) = \varphi (x - u t) \\ 
= \varphi (z) \, $ , $ z \equiv x - u t$ ; then from (\ref{eq: DS2}): 

\begin{equation}
 \varphi '' + u \, \varphi '  \, = \, - \frac{d}{d \varphi} F(\varphi)
\label{eq: phi}
\end{equation}

\noindent where $F$ is the primitive of $g$ , i.e. $g(\varphi) = dF / d \varphi$ . We may interpret (\ref{eq: phi}) 
as the equation of motion of a unit mass in a potential energy $F$ , subject to a frictional force 
proportional to the velocity. The dissipation parameter $u$ is the velocity of the kink. In Fig. \ref{fig:F_vs_phi_1} 
we plot the function $F$ obtained from integrating $g$ in (\ref{eq: g1}); the analytic expression, 
which involves the poly log function, is readily obtained with Mathematica. 

\noindent The kink solution displayed in Fig. \ref{fig:kink_1} corresponds, in terms of (\ref{eq: phi}), to the particle 
(of coordinate $\varphi$) starting with zero velocity at the maximum $\varphi = V_1$ and arriving 
(after an infinite time) at the secondary maximum $\varphi = V_3$ , also with zero velocity. The 
value of the dissipation parameter $u$ for which this is possible corresponds to the propagation velocity 
of the kink. Different velocities are possible transiently, for example, a kink initially steeper 
than the asymptotic shape will initially travel faster, and slow down as it attains the stable shape and velocity. 
This "shaping" of the signal expresses the existence of a stable, unique solitary wave solution.  
It motivated the electronic realization of an axon, and the corresponding influential dynamical system model, 
by Nagumo et al \cite{Nagumo1962}. 
Varying the clamp voltage $V_c$ modifies the potential $F$ , and the kink velocity $u$ 
changes correspondingly, as shown in Fig. 2. For increasing $V_c$ , the difference 
$F(V_1) - F(V_3)$ increases, while the secondary maximum at $V = V_3$ becomes less pronounced 
(Fig. \ref{fig:F_vs_phi_1}). Correspondingly, the kink velocity increases. At a critical clamp value 
$V_c \approx - 92.8 \, mV$ the secondary maximum disappears (the minimum at $V_2$ becomes an inflection point, 
then reverses curvature), so no kink solution exists for higher clamp voltages. 
Conversely, as $V_c$ is decreased, 
the difference $F(V_1) - F(V_3)$ decreases, goes through zero and becomes negative. Correspondingly 
the kink velocity also goes through zero and then reverses sign. In short, $F(V_1) - F(V_3)$ increases monotonically 
with increasing $V_c$ , as does the kink velocity $u$. There is a maximum positive velocity and a maximum 
negative velocity (the two are not the same). There is a particular clamp voltage ($V_c \approx 244.0 \, mV$ with 
our parameters) such that the kink is stationary ($u = 0$). Trivially, for each right-moving kink there is 
an identical mirror-image left-moving kink, if one inverts the boundary conditions at infinity. 
From Fig. \ref{fig:F_vs_phi_1} we also see that  
two more kink solutions exist, one connecting the maximum at $V_1$ with 
the minimum at $V_2$ (evidently travelling at a faster speed compared to the kink connecting $V_1$ 
and $V_3$), and a third one connecting $V_3$ and $V_2$. These solutions are linearly unstable, because 
the fixed point at $V_2$ is unstable; thus they would not be observed experimentally. However, they can still 
be "observed" numerically, as we see below. \\
It is interesting to put this problem in a "normal form", and see the connection to other kinks in condensed 
matter physics. The simplest function $F$ in (\ref{eq: phi}) which supports a kink solution of (\ref{eq: DS2}) 
has a maximum and a minimum, i.e. a cubic non-linearity. A kink solution exists connecting the maximum and the minimum, but it is unstable as the minimum is an unstable fixed point. The next simplest case is that $F$ has 
three extrema; assuming a single control parameter, we may write:  

\begin{equation}
 F(V) \,=\, a \, [2 (1 - \alpha) V^2 + \frac{4}{3} \alpha V^3 - V^4]
\label{eq: NF1}
\end{equation} 

\noindent $a > 0$ , $\alpha \le 1$ where we put one stable fixed point at 
$V_1 = 1$ and the unstable fixed point (the minimum of $F$) at $V_2 = 0$. The third (stable) fixed 
point is at $V_3 = (\alpha - 1)$. This is not the most general form: the choice $V_2 = 0$ forces $F$ to be 
an even function at the "coexistence point" $\alpha = 0$, as we discuss below; however, this choice allows 
to discuss unstable kink solutions also. Apart from this difference, this situation corresponds to (\ref{eq: g1}); 
the parameter $\alpha$ has the role of $V_c / V_N$, if $\chi_c / \chi$ is fixed. 
For $-1 < \alpha \le 1$ a stable kink with $V(x \rightarrow - \infty) = V_1$ and 
$V(x \rightarrow + \infty) = V_3$ exists, travelling with a speed $u$ which increases 
monotonically with increasing $\alpha$. The stationary kink is obtained for $\alpha = 0$; for 
$\alpha > 0$ the kink travels to the right and for $\alpha < 0$ to the left. The simplest stable kink 
is thus a solution of: 

\begin{equation}
\frac{\partial V(x, t)}{\partial t} - \frac{\partial^2 V}{\partial x^2} \, = \, 4 a \,  [ (1- \alpha ) V 
+ \alpha V^2 - V^3 ] 
\label{eq: DS3}
\end{equation}

\begin{figure}[H]
\includegraphics[width=\linewidth]{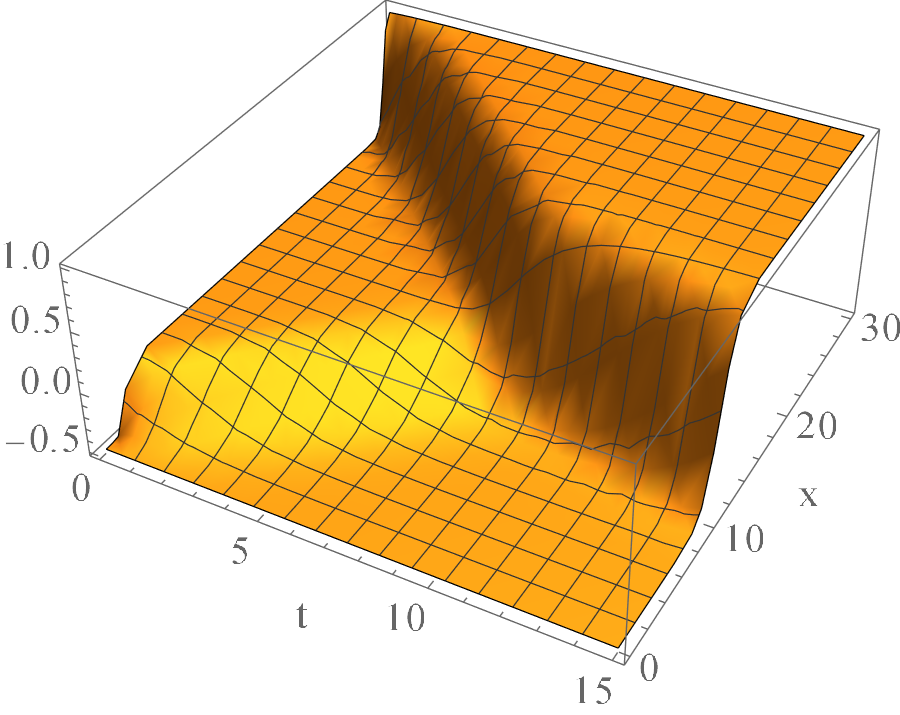}
\caption{A 3D plot showing the collision of two different kinks.  They are obtained integrating (\ref{eq: DS3}) 
with $a = 0.5$ , $\alpha = 0.5$, and appropriate initial conditions. Notice the velocity change after the collision. 
However, these kinks are linearly unstable and so would not be observed experimentally. }
\label{fig:Colli_1}
\end{figure}

\noindent The cubic nonlinearity is a feature of several reduced parameters models of nerve excitability, 
notably Fitzhugh's "BVP model" \cite{Fitzhugh}, and indeed of the original Van der Pol relaxation 
oscillator \cite{Van_der_Pol}, in appropriate coordinates. 
Two further kink solutions of (\ref{eq: DS3}) exist, connecting $V_1$ and $V_2$ , and $V_3$ and $V_2$. 
These are linearly unstable, but they can still be obtained numerically, with the trick of arranging 
for the unstable fixed point to be at $V = 0$, as we did in (\ref{eq: NF1}). In this way, one can 
even discuss collisions between different kinks: the only non-trivial example stemming from (\ref{eq: NF1}) 
is shown in Fig. \ref{fig:Colli_1}. 
 
\noindent Namely, the kink connecting $V_1$ and $V_2$ collides with the kink connecting $V_3$ and $V_2$ 
travelling in the opposite direction, resulting in the stable kink connecting $V_1$ and $V_3$ in the final state. \\
To ricapitulate: the fixed points of (\ref{eq: DS2}) are uniform, time-independent solutions which we 
might call "phases". Two fixed points can be connected by a kink. The fixed points are zeros of $g$, 
i.e. extrema of $F$, but the stable fixed points are maxima of $F$ while the unstable ones are minima. 
For the purpose of classifying, $F$ is analogous to minus the free energy of a Landau theory describing 
a corresponding phase transition. The stationary kink ($\alpha = 0$ in (\ref{eq: NF1})) is the interface 
separating two coexisting phases. For $\alpha \ne 0$ , one of the two phases is more stable and grows 
at the expense of the other (i.e. the kink moves). However, we must remember that our system is never 
in thermodynamic equilibrium. Even when the kink is stationary, there are macroscopic currents in 
the system (the clamp current and the channels current), and detailed balance in violated. The function 
$F$ derived from (\ref{eq: g1}), which is shown in Fig. \ref{fig:F_vs_phi_1} , has the same general form 
as (minus) the mean field free energy which describes the nematic - isotropic transition in liquid crystals 
\cite{Chaikin_Lub_Book}, or also the liquid - gas transition. For the former, and following the notation 
in \cite{Chaikin_Lub_Book}, the free energy $f$ as a function of the order parameter $S$ is: 

\begin{equation}
f \, = \, \frac{1}{2} a (T - T^*) S^2 - w S^3 + u S^4
\label{eq:NIT}
\end{equation}

\noindent where $S = P_2(cos \theta)$ , $P_2$ the Legendre Polynomial of order 2 and $\theta$ the angle between 
the molecular axis and the director vector. For fixed $V_c$ , the evolution of $- F$ for varying $\chi_c / \chi$ 
(where $F$ is the primitive of (\ref{eq: g1})) mirrors the evolution of (\ref{eq:NIT}) for varying temperature 
$T$. Namely, for small values of  $\chi_c / \chi$ there is a global minimum at positive $V$ (i.e. channels 
essentially open) and a secondary minimum at negative $V$ (channels essentially closed). Increasing 
$\chi_c / \chi$ one reaches a coexistence point where $-F$ has the same value at the two minima, after which 
the global minimum is at negative $V$ and the secondary minimum at positive $V$ 
(Fig. \ref{fig:F_vs_phi_1}), i.e. the stable 
phase is with channels essentially closed. As in (\ref{eq:NIT}) there are limits of meta-stability where the 
secondary minimum disappears. If we allow $V_c$ as a second control parameter, we find a coexistence 
line in the $V_c$ - $\chi_c / \chi$ plane ending in a critical point, i.e. the phenomenology of a liquid - gas 
transition. For parameter values on the coexistence line, the kink is stationary. \\ 
For the case of the stationary kink, one can write an implicit formula for the shape: with $u = 0$, 
multiplying (\ref{eq: phi}) by $\varphi ' $ and integrating from $- \infty$ to $x$ , with the boundary 
conditions  $\varphi ' \rightarrow 0$ , $\varphi \rightarrow \varphi_1$ for $x \rightarrow - \infty$ one finds 

\begin{equation}
\frac{d \varphi}{\sqrt{- F(\varphi) + F(\varphi_1)}} \, = \, - \sqrt{2} \, dx 
\label{eq:shape}
\end{equation} 
\\
\noindent For the stationary kink of (\ref{eq: DS3}), which occurs for $\alpha = 0$ , we have 
$F(\varphi) = a (2\varphi^2 - \varphi^4)$ , the maxima of $F$ are at $\varphi = \pm 1$ , and integrating 
(\ref{eq:shape}) we find $\varphi (x) = tanh (- \sqrt{2 a} \, x)$ . This is the same kink as in the mean field theory 
of the Ising ferromagnet, separating two domains of opposite magnetization \cite{Chaikin_Lub_Book}. 
It has a special symmetry (inversion about its center), stemming from the symmetry of this particular $F$ , 
which is an even function at the coexistence point $\alpha = 0$ . The function $F$ derived for the Artificial 
Axon from (\ref{eq: g1}) has no such symmetry, and correspondingly the stationary kink is not inversion symmetric 
about its center, as Fig. \ref{fig:kink_1} shows. For this kink too an analytic expression can be obtained 
from (\ref{eq:shape}) in terms of special functions. \\

\noindent {\it Conclusions.} We have discussed the occurrence of travelling kink solutions in a dynamical 
system which represents a space extended Artificial Axon. We considered the simplest limit: "fast" 
channels described by an equilibrium opening probability $P_O (V)$. Even so, the velocity of the kink 
represents a non trivial eigenvalue problem (\ref{eq: phi}). More generally, introducing 
channel dynamics increases the dimensionality of the dynamical system and leads to more structure 
(oscillations, limit cycles i.e. action potentials) as is well known. We point out a connection to similar kinks in other areas of condensed matter physics: 
some questions which can be asked of these systems are similar, for instance, effects beyond mean field \cite{Buijnsters2003, Buijnsters2014}. 
For us, this means replacing the uniform channel conductance with a space distribution of point - like channels, eventually interacting, eventually mobile. Introducing channel dynamics (see e.g. 
\cite{Morris_Lecar, Ziqi_1}), it may be interesting to extend this study to pattern formation 
in 2 space dimensions. In general, this system may inspire the construction of new 
reaction - diffusion systems \cite{Reac_diff_1} with interesting spatio - temporal dynamics.

\begin{acknowledgments}
\noindent This work was supported by NSF grant DMR - 1809381. 
\end{acknowledgments}

\bibliography{Xinyi_1_arch_refs}

\end{document}